\documentclass[epj]{svjour}
\usepackage{graphics,here}
\begin{document}
\title{Eta physics at threshold}
\author{ 
         P.~Moskal\inst{1,2}      \and
         H.-H.~Adam\inst{3}             \and
         A.~Budzanowski\inst{4}         \and
         R.~Czy{\.{z}}ykiewicz\inst{2}  \and
         D.~Grzonka\inst{1}             \and
         M.~Janusz\inst{2}              \and
         L.~Jarczyk\inst{2}             \and
         B.~Kamys\inst{2}               \and
         A.~Khoukaz\inst{3}             \and
         K.~Kilian\inst{1}              \and
         P.~Kowina\inst{1,6}      \and
         T.~Lister\inst{3}              \and
         W.~Oelert\inst{1}              \and
         T.~Ro{\.{z}}ek\inst{1,6} \and
         R.~Santo\inst{3}               \and
         G.~Schepers\inst{1}            \and
         T.~Sefzick\inst{1}             \and
         M.~Siemaszko\inst{6}           \and
         J.~Smyrski\inst{2}             \and
         S.~Steltenkamp\inst{3}         \and
         A.~Strza{\l}kowski\inst{2}     \and
         P.~Winter\inst{1}              \and
         M.~Wolke\inst{1}               \and
         P.~W{\"u}stner\inst{5}         \and
         W.~Zipper\inst{6}               
}                     
\offprints{}          
%
\institute{IKP, Forschungszentrum J\"{u}lich, D-52425 J\"{u}lich, Germany
     \and M.~Smoluchowski Institute of Physics, Jagellonian University, PL-30-059 Cracow, Poland
     \and IKP, Westf\"{a}lische Wilhelms--Universit\"{a}t, D-48149 M\"{u}nster, Germany
     \and Institute of Nuclear Physics, PL-31-342 Cracow, Poland 
     \and ZEL,  Forschungszentrum J\"{u}lich, D-52425 J\"{u}lich,  Germany
     \and Institute of Physics, University of Silesia, PL-40-007 Katowice, Poland}
\date{Received: date / Revised version: date}
%
\abstract{
  The production of $\eta$ and $\eta^{\prime}$
  mesons in elementary nucleon-nucleon collisions has been
  investigated at the synchrotrons CELSIUS, COSY and SATURNE.
  The high quality proton beam  with low emittance
  and small momentum spread permits to study the
  creation of these mesons very close to the kinematical threshold,
  where --~due to the rapid growth of the phase space volume~--
  the total cross section
  increases by orders of magnitude
  over a few MeV range of the excess energy.
  The magnitude and energy dependence of the total cross section as well as  the 
  occupation distribution of the phase space serve as observables for  investigating
  the mechanisms underlying the production processes and the interaction 
  of mesons with nucleons. The precise data on the $\eta$ and $\eta^{\prime}$ creation 
  via the $pp\to pp \eta(\eta^{\prime})$ reactions allowed to settle the general features
  of the $\eta$ and $\eta^{\prime}$ meson production and revealed the sensitivity of the 
  mentioned observables to 
  the nucleon-nucleon-meson final state interaction. 
  The particular production properties, like for example the determination 
  of the dominating exchange processes  which lead to the excitation of the 
  S$_{11}$ nucleon isobar in the case of $\eta$ creation,
  must be established by  confrontation 
  with other observables. 
  The present status of this investigation with an emphasis on the results
  of the COSY-11 collaboration is briefly presented.
  The available data
  are interpreted 
  in view of the production mechanism and the meson-nucleon interaction.
} 
\maketitle
\section{Manifestation of the $\eta$-nucleon-nucleon interaction}
\label{intro}
\begin{figure}[H]
\centerline{
\resizebox{0.34\textwidth}{!}{%
  \includegraphics{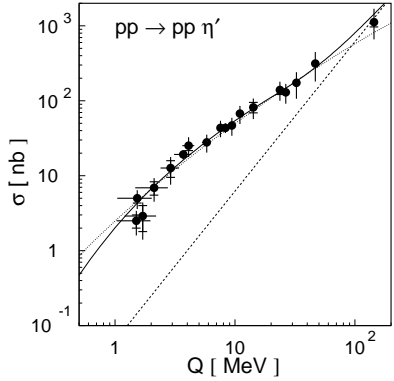}
}
}
  \vspace{-0.3cm}
  \caption{ \label{cross_etap}
      Total cross section for the 
      $pp \rightarrow pp \eta^{\prime}$ reaction as a function of the centre--of--mass
      excess energy Q. Data are from
      refs.~\cite{etap_data,hibou}.
      The solid line
      shows the phase--space distribution with inclusion of proton--proton strong
      and Coulomb interactions. The dotted line indicates the parametrization of
      reference~\cite{faldt209}
      with $\epsilon = 0.3$ and the dashed line
      indicates a phase--space integral normalized arbitrarily.
      Figure and title are adapted from reference~\cite{review}.
  }
\end{figure}

\begin{figure}[h]
\centerline{
\resizebox{0.3\textwidth}{!}{%
  \includegraphics{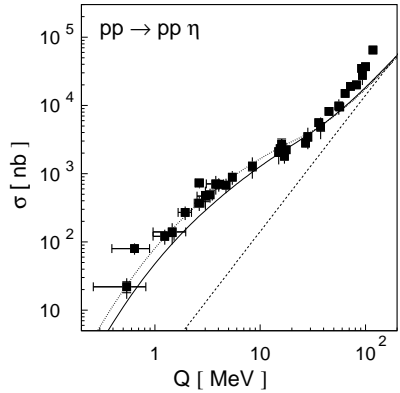}
}
}
  \caption{ \label{cross_eta}
    Total cross section for the $pp \rightarrow
    pp \eta$ reaction as a function of the centre--of--mass excess energy Q. Data
    are from refs.~\cite{hibou,eta_data}.
    The dashed line indicates a phase--space integral normalized
    arbitrarily. The phase--space distribution with inclusion of proton--proton
    strong and Coulomb interactions fitted to the data in the excess energy range
    between 15 and $40\,\mbox{MeV}$ is shown as the solid line. Additional
    inclusion of the proton--$\eta$ interaction is indicated by the dotted line.
    The scattering length $a_{p\eta} = 0.7\,\mbox{fm} + i\,0.4\,\mbox{fm}$ and the
    effective range parameter $b_{p\eta} = -1.50\,\mbox{fm} -
    i\,0.24\,\mbox{fm}$~\cite{greenR2167} have been arbitrarily chosen.
    The figure and caption are adapted from reference~\cite{review}.
  }
\end{figure}

\begin{figure}[H]
\centerline{
\resizebox{0.45\textwidth}{!}{%
  \includegraphics{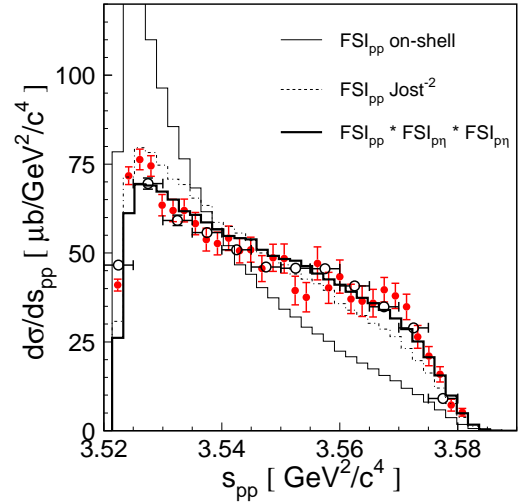}
}
}
  \vspace{-0.3cm}
  \caption{ \label{dsigmapodspp}
          Experimental distribution of the square of the proton-proton invariant mass ($s_{pp}$)
          determined experimentally
          for the $pp  \to pp\eta$ reaction at the excess energy of Q~=~15.5~MeV
          by the collaborations COSY-11~\cite{meson} (closed circles)
          and TOF~\cite{TOFeta} (open circles).  The TOF data have been normalized 
          to give the same total cross section as those of the COSY-11.
          The integral of the phase space
          weighted by the proton-proton on-shell scattering amplitude 
          and by the inverse of the proton-proton squared Jost function~\cite{niskanen107}
          is represented by the thin solid and dashed line, respectively.
          The thick solid line shows the result of the calculations where 
          the occupation density of the phase space 
          was weighted by the inverse of the Jost function
          multiplied by the square of the proton-$\eta$ scattering amplitude.
          The latter was calculated  taking arbitrarily the scattering length 
          $a_{p\eta} = 0.7\,\mbox{fm} + i\,0.4\,\mbox{fm}$.
          The lines have been normalized in amplitude to the data.
  }
\end{figure}

  In the last decade  large experimental as well as theoretical efforts
  were concentrated on the study of the creation of 
  $\eta$ and $\eta^{\prime}$ mesons
  via the hadronic interactions. Measurements of the production of these mesons
  in the elementary nucleon-nucleon collision have been performed 
  in the vicinity of the kinematical threshold where only one partial wave in both
  initial and final state
  is expected to contribute to the production process. 
  For example in case of the proton-proton collision the dominance of the $^3P_{0}\to Ss$ transition
  is expected up to an excess energy 
  of about 40~MeV and 100~MeV for $\eta$ and $\eta^{\prime}$ meson,
  respectively~\cite{review}. This simplifies significantly the 
  interpretation of the data, yet still appears to be challenging
  due to the three particle final state system with a complex hadronic potential.
  The determined energy dependences of the total cross section 
  for $\eta^{\prime}$~\cite{etap_data,hibou} and $\eta$~\cite{hibou,eta_data}
  mesons in  proton-proton collisions
  are presented in figures~\ref{cross_etap}~and~\ref{cross_eta}.
  Comparing the data to the arbitrarily normalized phase-space integral (dashed lines)
  reveals that the proton-proton FSI enhanced the total cross section by more than an order
  of magnitude for low excess energies. One recognizes also that in the case of the $\eta^{\prime}$
  meson the calculation 
    --assuming that the on-shell proton-proton amplitude
    exclusively determines the phase-space population--
  describes the data very well (solid line).
  This indicates that the  proton-$\eta^{\prime}$ interaction is too small to be observed
  within the present accuracy~\cite{swave}.
  In the case of the $\eta$ meson the increase of the
  total cross section for very low and very high energies is much larger than expected 
  from the  final state interaction between protons. The excess at higher energies
  can be assigned to the onset of higher partial waves, and 
  the enhancement at threshold can be plausibly explained by the influence of the attractive 
  interaction between the $\eta$ meson and the proton. \\
  Though the simple phenomenological treatment --based on factorization of the transition amplitude
  into the constant primary production and the on-shell incoherent pairwaise interaction
  among particles-- works well for the energy dependence  of the total cross section,  
  it fails completely as far as the description of the differential cross section 
  is concerned (see the thin solid line in figure~\ref{dsigmapodspp}). 

\vspace{-0.8cm}

\vspace{-1.2cm}
  This discrepancy is rather too large to be utterly caused 
  by  the underestimation of the S-wave proton-$\eta$ interaction.
  An explanation could be 
  a contribution from P-wave proton-proton interaction~\cite{hanhart}
  or 
  a significant influence of the off-shell 
  effects of the interaction between outgoing particles.
  Indeed, a much better description of the differantial 
  cross section shown in figure~\ref{dsigmapodspp}
  is achieved when
  the enhancement due to the proton-proton 
  FSI is modeled by the inverse of the squared Jost function instead of the 
  on-shell amplitude of the elastic proton-proton scattering 
  (see thick solid line in fig.~\ref{dsigmapodspp}).
  The Jost function accounts 
  for the off-shell effects of the interacting particles, however
  it gives wrong energy dependence of the total cross section, and 
  unfortunately 
  it depends rather strongly on the potential
  model~\cite{baru,review} and applied formalism~\cite{kle01}.
  Therefore in the framework of this 
  phenomenology it is not possible --with a satisfactory accuracy-- 
  to separate the contribution of the
  proton-$\eta$ interaction from other effects 
  in a model independent way. Thus, more
  sophisticated theoretical calculations are required.
  An estimation of the nucleon-$\eta$
  scattering parameters has successfuly been performed by comparing the 
  close-to-threshold data of the $\eta$ meson photoproduction 
  on a deuteron~\cite{hejny} to the calculations performed 
  to the first order rescattering in the two-body subsystems~\cite{elster}.
  The recent investigation of Fix and Arenh{\"o}vel~\cite{fix} shows, however,
  that the three-body dynamics contribute substantially to the 
  total and differential cross sections of the $\eta$ photoproduction on the deuteron,
  and thus it cannot be neglected in the quantitative analysis.

\vspace{-0.8cm}
  Data of the $\eta$-NN system
   created 
  via the  hadronic interaction of nucleons still awaits a strict theoretical explanation.  
  The hitherto performed analysis of the
  energy dependence of the total cross section reveals the sensitivity of the data to the 
  $\eta$-nucleon interaction.
  Clearly, further comparative studies of the $\eta$-nucleon-nucleon  system created in photo-
  and hadro-production should lead to a better determination of the $\eta$-nucleon potential 
  and specifically should be helpful in distinguishing  between effects 
  due to the production dynamics and the final state interaction among the produced
  particles.

\section{Production dynamics}
  It is rather well established that close to threshold the energy dependence
  of the total cross section is due to the interaction among the outgoing particles
  and that the entire production dynamics manifests itself only in a single constant
  which determines the magnitude of the total cross section~\cite{review}.
  Therefore in spite of the precise measurements of the  total cross section 
  for the creation of $\eta$ and $\eta^{\prime}$ mesons in  proton-proton-- 
  as well as of $\eta$ meson in  proton-neutron collisions
  there is still a lot of ambiguities in the description of the 
  mechanism underlying the production process.  
  It is generally anticipated~\cite{eta_theo,eta_rho,eta_all} that the $\eta$ meson is produced
  predominantly via the excitation of the S$_{11}$ baryonic resonance $N^{*}(1535)$,
  whose creation is induced through the exchange of the virtual $\pi$, $\eta$, $\rho$,
  $\sigma$ and $\omega$ mesons, however at present it is still not established what are 
  the relative contributions originating 
  from a particular meson. Measurements of the total cross
  section in  different isospin channels put some more limitations to the models,
  yet still the $\eta$ meson production in the $pp\to pp\eta$ and $pn\to pn\eta$
  can be equally well described by eg. assuming the $\rho$ meson exchange dominance~\cite{eta_rho}
  or by taking contributions from the pseudoscalar and vector meson exchanges~\cite{eta_all}.
  Therefore for a full understanding of the production dynamics the determination  of 
  polarisation observables is mandatory.  Already the precise measurement of the
  beam analyzing power could  exclude one of the above mentioned possibilities.
  The first measurement of that quantity have been recently performed~\cite{poleta},
  but for a conclusive inference a better accuracy of the data is required.
  
  Until now it was also not possible to determine unambiguously the mechanism of the 
  production of the $\eta^{\prime}$ meson~\cite{review}.  Model uncertainties and 
  in practice only one number --namely the magnitude of the total cross section 
  of  the $pp \to pp\eta^{\prime}$ reaction-- serving as input for theory 
  are by far not sufficient to distinguish between the mesonic, nucleonic 
  or resonant production currents~\cite{etap_theo}. 
  The understanding of that mechanism
  on the hadron and quark-gluon level will shed a light on the structure of the
  $\eta^{\prime}$ meson.
  A possibly large glue content of the $\eta^{\prime}$ wave function
  and the dominant
  flavour--singlet combination of its quarks component may cause that the
  dynamics of its production process in nucleon--nucleon collisions is
  significantly different from that responsible for the production of other
  mesons.  A creation of that meson via a fusion of gluons excited in the 
  interaction region should lead to the same production yield in 
  proton-proton and proton-neutron collisions because gluons do not distinguish
  between flavours~\cite{bass}. 
  Therefore it is interesting to determine experimentally the 
  ratio 
  $R_{\,\eta^{\prime}} = \sigma(pn \rightarrow pn \eta^{\prime})/
  \sigma(pp \rightarrow pp \eta^{\prime})$  and to compare it to the already known
  value of $R_{\eta}~\approx~6.5$~\cite{calen2667}.  
  In the extreme case when the $\eta^{\prime}$ meson is created only from glue
  excited in the interaction region the ratio $R_{\,\eta^{\prime}}$ should be close
  to unity after the correction for the final and initial state interactions 
  between the interacting nucleons~\cite{bass}.
  On the contrary a production mechanism dominated by 
  the isovector meson exchange would yield a significantly larger ratio as already
  observed in the case of the $\eta$ meson.
  The appropriate experiments are in preparation 
  at the COSY-11 facility~\cite{pnetapc11} and the first feasibility study has already 
  been  accomplished sucessfully.
\vspace{-0.5cm}
\section*{Acknowledgement}
\vspace{-0.3cm}
{\small
 The work has been partly supported by the European Community - Access to
Research Infrastructure action of the Improving Human Potential Programme
}

\end{document}